**Title:** International trade of fruits between Portugal and the world

**Author:** Vítor João Pereira Domingues Martinho

Unidade de I&D do Instituto Politécnico de Viseu

Av. Cor. José Maria Vale de Andrade

Campus Politécnico

3504 - 510 Viseu

PORTUGAL

e-mail: vdmartinho@esav.ipv.pt

# International trade of fruits between Portugal and the world

**Abstract**

For Portugal there are few or none works about the international trade of fruits between Portugal and the other countries. In this work it aims to analyze the more recent data for the Portuguese international trade of fruits. They were used data for the years from 2006 to 2010, available by the INE (Statistics Portugal), gently given by the AICEP (Trade & Investment Agency). To complement this data analysis they were made some estimations with several econometrics method and based in the neoclassical theory, with the absolute convergence model. It was concluded that the biggest relationship, in the international trade of fruits, is with the European countries and there are not statistical regularity in the estimations and the data are not stationary.

**Keyword:** Fruits, international trade, data analysis.

## 1. Introduction

Portugal has excellent conditions to produce fruits, because has a climate very favorable for these productions. However, this is a sector with some problems, because is much deregulated economic activity as result of the common agricultural policy (CAP). The CAP is little focused for the south countries of the Europe, because this, some authors say that the CAP is economically inefficient and socially unjust. Economically inefficient, because induce the farmers to decide for productions with more subsidies and for productions more adjusted to the local conditions and socially unjust, because is a policy that support the biggest farmers. So, only from here there is a long way to go.

In the recent years the different reforms of the CAP tried to solve some of these situations, but are not enough. One of this trying is the suspension of the CAP payments from the production, with the objective to guide production management and resources distribution to be in connection only with the market prices and structural capacities. The results demonstrate which the payments of CAP not connected with the production have significant economic consequences and the expected augments in the prices do not balance the failure of the Agenda 2000 area payments (Fragoso et al., 2009).

Anyway the international trade of fruits is an important business area, however some countries have comparative advantages. For example, in the ASEAN countries (Philippines, Indonesia, Singapore and Thailand), Singapore has structural advantage in 5 products (ground-nuts, hazelnuts, plums, apricots and walnuts), Philippines has structural advantage in 3 products (tomatoes prepared or preserved, tomatoes whole or in pieces and cherries), Thailand and Malaysia have structural advantage in 2 products, while Indonesia has structural advantage in 1 product (cashew nuts). Malaysia has structural advantage only in tomatoes and apple juice but structural disadvantage in other products such as cashew nuts, walnuts and fruit and vegetable juice (Emmy and Ismail, 2009).

For the NAFTA countries, namely for USA and Mexico, some studies analyze the consequences of this economic integration in the international trade of fruits and vegetables. The conclusion is that the import price elasticities show which imports are not susceptible to price adjusts. Nevertheless, the income elasticities of import demand differ by products. There is trade growth as consequence of the NAFA in the vegetable and fruit trade. The amount of trade creation is larger than the quantity of trade diversion in most products analyzed (Karemera et al., 2007).

In the European Union the import regimes after the Uruguay Round are based on ingress prices that in practice function like lowest prices. On other hand, in this time, the European Union has celebrated trade favorite agreements namely with Southern Mediterranean countries which are significant suppliers of fruit and vegetables to the European Union. In the export side, the subsidies do not look capable to increase the exports of eligible products. Consequently, the European Union must consider if should maintain those subsidies (Cioffi and dell´Aquila, 2004). The models illustrate which prices work in a different way when import prices are different of the start entry price (Cioffi et al., 2010). Trade openness has a great effect on European fruit sector, at productive and commercial level. European Union fruit sector, at productive and export level, are expected to decrease considerably. European vegetables production and exports are reasonably protected and are expected to earning from the diminution in European Union fruit sector (Bunte, 2005).

Analyzing the international trade of fruits between the South Mediterranean Countries and the European Union, the Magreb region, achieve comparatively poorer than the exporting countries from in the interior of the European. One explanation for this is the trade variation effects of European Union integration and the increasing demands for quality and service forced by horticultural retailers, which are improved in developed countries. This could too clarify why Israel functions well than exports flows from the Magreb and Mashrek subregions (Coque and Selva, 2007).

In Africa, agricultural trade between the countries of ECOWAS (a group of 15 countries of the West Africa which has eliminated tariffs on agricultural trade between each one) is superior than the expected. This does not signify that there are no non-tariff barriers inside ECOWAS, but it implies that any barriers are less damaging to agricultural trade in ECOWAS than in the rest of the world. This shows that African countries are not reluctant to agricultural trade, and local operators have been successful at finding trade new destinies (Seck et al., 2010).

## 2. Data analysis

Observing the table 1 below, Portugal import, from South Africa, specifically citrus fruit, grapes, apples, pears and quinces. From Angola and Cape Verde do not import any fruits. Brazil export to Portugal, namely, dates, figs, pineapples or pineapple, mangoes, mangosteens, fresh or dried and melons, watermelons and papaws (papayas), fresh, as expected because is a tropical country. Costa Rica export to Portugal, namely, bananas and dates, figs, pineapples or pineapple, mangoes, mangosteens, fresh or dried. From the United States the Portuguese import, specifically, other nuts, fresh or dried, whether or not shelled or peeled. From China Portugal import several fruits, without regularity. Turkey send to Portugal, namely, dates, figs, pineapples or pineapple, mangoes, mangosteens, fresh or dried and grapes. India send to the Portuguese coconuts, Brazil nuts from Brazil and cashew nuts, fresh or dried, etc and other nuts, fresh or dried, whether or not shelled or peeled. From Germany and France import apples, pears and quinces, fresh. From Belgium import dates, figs, pineapples or pineapple, mangoes, mangosteens, fresh or dried. From Spain, Estonian, Ireland, Italy, Luxembourg and Poland import several fruits without a visible majority. From Greece

import fruits, cooked or not, frozen, containing added sugar or sweetener. Holland send to Portugal, namely, other fresh fruit and the United Kingdom send bananas, including plantains (platains), fresh or dried and dates, figs, pineapples or pineapple, mangoes, mangosteens, fresh or dried.

**Table 1. Fruits, in different forms, import percentage relatively to the total of each country**

| | Year | South Africa | Angola | Cape Verde | Brazil | Costa Rica | United States of America | China | Turkey | India | Germany | Belgium | Spain | Estonian | France | Greece | Holland | Ireland | Italy | Luxembourg | Poland | United Kingdom |
|---|---|---|---|---|---|---|---|---|---|---|---|---|---|---|---|---|---|---|---|---|---|---|
| Coconuts, Brazil nuts from Brazil and cashew nuts, fresh or dried, | 2006 | | | | 6 | | | | | 60 | 0 | | 0 | | 0 | | 1 | 0 | | | | 3 |
| | 2007 | | | | 5 | 0 | | | | 58 | 0 | | 0 | | 0 | | 2 | | | | | 2 |
| | 2008 | | | | 5 | 0 | 0 | 0 | | 69 | 1 | | 0 | | 0 | | 2 | | | | | 0 |
| | 2009 | | | | 8 | | | 0 | 0 | 30 | 5 | | 0 | | 0 | | 2 | | | | 3 | 8 |
| | 2010 | | | | 7 | 0 | | 0 | | | 2 | | 0 | | 0 | | 4 | | | | | 11 |
| Other nuts, fresh or dried, whether or not shelled or peeled | 2006 | | | | | | 99 | 84 | 7 | 38 | 3 | | 6 | | 11 | 32 | 8 | | 2 | | | 2 |
| | 2007 | | | | 0 | | 92 | 2 | 11 | 41 | 4 | 0 | 5 | | 9 | 19 | 3 | | 2 | | | 1 |
| | 2008 | | | | | | 99 | 6 | 13 | 30 | 15 | 0 | 5 | 2 | 13 | | 5 | | 1 | | | 1 |
| | 2009 | | | | | | 100 | 5 | 8 | 68 | 12 | | 6 | 4 | 14 | | 4 | | 1 | | 13 | 0 |
| | 2010 | | | | | | 100 | 3 | 14 | 98 | 16 | 2 | 6 | 4 | 14 | | 8 | | 1 | | | 0 |
| Bananas, including plantains (platains), fresh or dried | 2006 | | | | 3 | 42 | | | | | 1 | 47 | 9 | | 22 | | 3 | | 63 | | | 15 |
| | 2007 | | | | 3 | 55 | | | | | 0 | 3 | 10 | | 16 | | 3 | | 35 | | | 12 |
| | 2008 | | | | 3 | 41 | 0 | | | | 0 | 5 | 13 | 4 | 14 | | 0 | | 4 | | | 11 |
| | 2009 | | | | 1 | 40 | | | | | 1 | 2 | 10 | | 25 | | 4 | 60 | 1 | | | 64 |
| | 2010 | | | | 0 | 32 | | | | | 0 | 0 | 12 | | 14 | | 0 | | 0 | | | 36 |
| Dates, figs, pineapples or pineapple, mangoes, mangosteens, | 2006 | 1 | | | 48 | 57 | 0 | | 34 | 1 | 1 | 30 | 8 | | 2 | | 23 | | 2 | | | 57 |
| | 2007 | 7 | | | 38 | 44 | 2 | 1 | 37 | 1 | 1 | 41 | 9 | | 3 | | 22 | | 3 | | | 49 |
| | 2008 | 8 | | | 40 | 59 | 0 | | 37 | 1 | 3 | 39 | 10 | 4 | 2 | | 17 | | 1 | | | 17 |
| | 2009 | 5 | | | 34 | 59 | 0 | | 34 | 1 | 5 | 29 | 11 | 9 | 2 | | 9 | 40 | 0 | 62 | | 12 |
| | 2010 | 0 | | | 34 | 67 | 0 | 0 | 26 | 2 | 5 | 17 | 9 | 4 | 2 | | 12 | | 0 | 15 | | 15 |
| Citrus fruit, fresh or dried | 2006 | 8 | | | 7 | | | 0 | | | 11 | 1 | 7 | | 0 | | 2 | | 0 | | | 0 |
| | 2007 | 40 | | | 5 | | | | | | 12 | 0 | 8 | | 1 | | 5 | | 4 | | | 16 |
| | 2008 | 39 | | | 6 | | | | | | 1 | 0 | 8 | 13 | 1 | | 5 | | 14 | | | 27 |
| | 2009 | 31 | | | 7 | | | | | | 4 | 4 | 9 | 0 | 1 | | 8 | | 19 | | | |
| | 2010 | 80 | | | 8 | | | | | | 3 | 3 | 10 | 0 | 0 | | 8 | | 2 | | | |
| Grapes, fresh or dried | 2006 | 57 | | | 0 | | 1 | | 42 | | 7 | 7 | 11 | 100 | 0 | | 6 | | 3 | | | 0 |
| | 2007 | 20 | | | 0 | | 0 | | 35 | | 7 | 2 | 12 | | 0 | 7 | 6 | | 12 | | 29 | 0 |
| | 2008 | 22 | | | | | 0 | | 34 | | 15 | 3 | 10 | 4 | 2 | 7 | 8 | | 18 | | | 21 |
| | 2009 | 22 | | | | | 0 | 30 | 39 | | 9 | 3 | 11 | 23 | 6 | 3 | 5 | | 16 | | | 2 |
| | 2010 | 9 | | | | | 0 | 23 | 38 | | 13 | 2 | 9 | 12 | 5 | | 4 | | 27 | | | 0 |
| Melons, watermelons and papaws (papayas), | 2006 | | | | 29 | 1 | 0 | | | | 4 | | 13 | | 0 | | 2 | | 0 | | | |
| | 2007 | | | | 30 | 1 | 4 | | | | 6 | | 13 | | 0 | 3 | 1 | | | | | 3 |
| | 2008 | | | | 32 | 0 | 0 | | | | 15 | | 13 | 31 | 0 | 15 | 3 | | | | | 6 |
| | 2009 | 0 | | | 34 | 1 | | | | | 15 | | 13 | 7 | 0 | 16 | 0 | | 38 | | | |
| | 2010 | | | | 33 | 1 | | | | | 15 | | 13 | 26 | 0 | 2 | 3 | | 0 | 56 | | 6 |
| Apples, pears and quinces, fresh | 2006 | 18 | | | 7 | | | | | | 42 | 2 | 10 | | 54 | | 12 | | 6 | | | |
| | 2007 | 26 | | | 18 | | | 62 | | | 37 | 6 | 10 | | 63 | | 19 | | 24 | | | 13 |
| | 2008 | 20 | | | 13 | | | 83 | | | 24 | 10 | 9 | 10 | 59 | | 4 | | 30 | | 100 | 9 |
| | 2009 | 31 | | | 16 | | | 14 | | | 6 | 7 | 8 | 28 | 32 | | 12 | | 39 | | 68 | 10 |
| | 2010 | 9 | | | 16 | | | 2 | | | 2 | 3 | 6 | 20 | 52 | | 10 | | 44 | | | 23 |
| Apricots, cherries, peaches (including nectarines), | 2006 | 14 | | | | | | | | | 17 | 1 | 20 | | 1 | | 2 | | 0 | | | |
| | 2007 | 7 | | | | | | | | | 17 | 1 | 16 | | 1 | | 2 | | | | | |
| | 2008 | 12 | | | | | | | | | 10 | 2 | 15 | 28 | 1 | | 0 | | 1 | | | 1 |
| | 2009 | 10 | | | | | | | 2 | | 11 | 8 | 15 | 4 | 0 | | 1 | | 0 | | | |
| | 2010 | 2 | | | | | | | | | 7 | 4 | 18 | 22 | 0 | 3 | 1 | | 0 | | | 1 |
| Other fresh fruit | 2006 | | | | 1 | | | | | 0 | 9 | 7 | 14 | | 3 | | 35 | | 23 | | | 20 |
| | 2007 | | | | 1 | | 0 | | | 0 | 7 | 24 | 14 | 100 | 3 | | 30 | | 20 | | | 1 |
| | 2008 | 0 | | | 1 | | | 0 | | 0 | 5 | 28 | 14 | 4 | 4 | 3 | 34 | | 28 | | | 2 |
| | 2009 | 0 | | | 1 | 0 | | | | | 16 | 24 | 13 | 25 | 16 | 7 | 40 | | 22 | | | 0 |
| | 2010 | 0 | | | 1 | | | | | 0 | 19 | 54 | 14 | 12 | 7 | | 39 | | 21 | | | 3 |
| Fruits, cooked or not, frozen, containing added sugar or sweetener | 2006 | | | | | | | 8 | 1 | | 0 | 6 | 0 | | 3 | 68 | 5 | | | | 100 | 0 |
| | 2007 | | | | | | | 31 | | | 0 | 22 | 1 | | 2 | 71 | 7 | | 0 | | 71 | 0 |
| | 2008 | | | | | | | 3 | 1 | | 1 | 12 | 0 | | 1 | 75 | 22 | | 2 | | | 2 |
| | 2009 | | | | | | | 19 | 4 | | 1 | 24 | 1 | | 2 | 74 | 13 | | 2 | | | 0 |
| | 2010 | | | | 2 | | 0 | 50 | | | 1 | 14 | 0 | | 4 | 95 | 12 | | 4 | 29 | | 0 |
| Fruit provisionally preserved but unsuitable in that state | 2006 | | | | | | | | | | 3 | 0 | 1 | | 0 | | 0 | | | | | 1 |
| | 2007 | | | | | | | | | | 4 | | 1 | | 1 | | 0 | | | | | 1 |
| | 2008 | | | | | | | | | | 3 | 0 | 2 | | 1 | | 0 | | 0 | | | 0 |
| | 2009 | | | | | | | | | 0 | 8 | | 2 | | 1 | | 0 | | | | | |
| | 2010 | | | | | | | | | | 6 | | 1 | | 1 | | 1 | | | | | |
| Dried fruit, mixtures thereof or nuts | 2006 | 1 | | | 0 | | 0 | 8 | 16 | 0 | 2 | 0 | 1 | | 2 | | 0 | | 0 | | | 2 |
| | 2007 | 0 | | | 0 | | 1 | 4 | 17 | 0 | 3 | 0 | 1 | | 1 | | 0 | | 0 | | | 0 |
| | 2008 | 0 | | | 0 | | | 8 | 15 | | 7 | 1 | 1 | | 1 | | 0 | | 0 | | | 2 |
| | 2009 | 0 | | | 0 | | 0 | 33 | 13 | 0 | 7 | | 1 | | 0 | | 1 | | 0 | | 17 | 3 |
| | 2010 | | | | 0 | | 0 | 21 | 22 | | 12 | | 1 | | 0 | | 1 | | 0 | | | 3 |
| Peel of citrus fruits, melons and / or melons, fresh, dried or frozen, etc. | 2006 | | | | | | | | | | 0 | | 0 | | | | | | | | | |
| | 2007 | | | | | | | | | | 0 | | 0 | | | | | | | | | |
| | 2008 | | | | | | 0 | 0 | | | 0 | | 0 | | 0 | | | | | | | 0 |
| | 2009 | | | | | | | | | | 0 | | 0 | | 0 | | 0 | | | | | |
| | 2010 | | | | | | | | | | | | 0 | | 0 | | | | | | | |

Portugal export (table 2) to South Africa, namely, dates, figs, pineapples or pineapple, mangoes, mangosteens, fresh or dried, to Angola other nuts, fresh or dried, whether or not shelled or peeled, to Cape Verde and Brazil apples, pears and quinces, fresh, to United States other nuts, fresh or dried, whether or not shelled or peeled. To Germany, France, Ireland, Poland and United Kingdom export, specifically, apples, pears and quinces, fresh. To Holland export other fresh fruits, to Italy bananas, including plantains (platains), fresh or dried and to Luxembourg other nuts, fresh or dried, whether or not shelled or peeled.

**Table 2. Fruits, in different forms, export percentage relatively to the total of each country**

| | Year | South Africa | Angola | Cape Verde | Brazil | Costa Rica | United States of America | China | Turkey | India | Germany | Belgium | Spain | Estonian | France | Greece | Holland | Ireland | Italy | Luxembourg | Poland | United Kingdom |
|---|---|---|---|---|---|---|---|---|---|---|---|---|---|---|---|---|---|---|---|---|---|---|
| Coconuts, Brazil nuts from Brazil and cashew nuts, fresh or dried, | 2006 | | 10 | 4 | | | 4 | | | | 0 | 0 | 0 | | 0 | | 0 | 0 | 0 | 0 | | 0 |
| | 2007 | | 4 | 7 | | | 6 | | | | 0 | | 0 | | 0 | | 0 | 0 | 0 | 5 | | 0 |
| | 2008 | 0 | 5 | 5 | | | 5 | | | | 0 | 0 | 0 | | 0 | | 0 | 0 | 0 | 0 | | 0 |
| | 2009 | 0 | 7 | 4 | | | 5 | | | | 0 | 0 | 0 | | 0 | | | 0 | | 0 | | 0 |
| | 2010 | | 7 | 2 | | | 4 | 100 | | | 0 | 0 | 0 | | 0 | | | 0 | 0 | 0 | | 0 |
| Other nuts, fresh or dried, whether or not shelled or peeled | 2006 | 11 | 51 | 2 | 32 | | 76 | | | | 13 | 33 | 25 | | 11 | | 2 | 0 | 22 | 98 | | 3 |
| | 2007 | 1 | 48 | 2 | 37 | | 72 | | | | 33 | 29 | 20 | | 9 | 100 | 1 | 0 | 26 | 21 | 0 | 3 |
| | 2008 | 2 | 48 | 1 | 34 | | 70 | | | 100 | 11 | 17 | 16 | | 15 | | 1 | 0 | 7 | 26 | | 0 |
| | 2009 | 5 | 43 | 1 | 24 | | 67 | | | | 7 | 76 | 14 | | 19 | | 0 | | 17 | 58 | | 1 |
| | 2010 | 56 | 36 | 2 | 20 | | 58 | | | | 16 | 71 | 14 | | 14 | | 2 | 0 | 23 | 30 | | 3 |
| Bananas, including plantains (platains), fresh or dried | 2006 | | 0 | 0 | | | | | | | 0 | 5 | 13 | | 19 | | | 0 | 60 | | | 11 |
| | 2007 | | | 0 | 0 | | | | | | 0 | 6 | 16 | | 10 | | | | 48 | | 4 | 9 |
| | 2008 | | | 0 | 0 | | | | | | 1 | | 17 | | | | | | 61 | 2 | | 0 |
| | 2009 | | | | 0 | | | | | | | | 14 | | 0 | | | | 56 | 5 | | |
| | 2010 | | | 0 | 0 | | | | | | | | 10 | | 0 | | 0 | | 39 | 11 | | |
| Dates, figs, pineapples or pineapple, mangoes, mangosteens, | 2006 | | 7 | 5 | | | 14 | | | | 18 | 0 | 18 | | 0 | | 2 | 0 | 13 | 0 | | 0 |
| | 2007 | 99 | 9 | 6 | | | 11 | | | | 1 | 0 | 13 | | 1 | | 3 | | 19 | 26 | | 0 |
| | 2008 | 98 | 11 | 7 | | | 11 | | 100 | | 6 | 0 | 15 | | 2 | 100 | 3 | 0 | 26 | 20 | | 0 |
| | 2009 | 95 | 10 | 6 | | | 9 | | | | 0 | 0 | 16 | | 1 | 100 | 1 | 0 | 23 | 0 | 6 | 0 |
| | 2010 | 44 | 7 | 8 | | | 9 | | | | 0 | 0 | 14 | | 0 | | 1 | | 35 | 0 | | 0 |
| Citrus fruit, fresh or dried | 2006 | | 0 | 24 | | | | | | | 4 | | 15 | | 6 | | | | 2 | 0 | 29 | 2 |
| | 2007 | | 1 | 24 | | | | | | | 4 | 8 | 24 | | 8 | | 0 | | 3 | | 39 | 1 |
| | 2008 | | 1 | 25 | | | | | | | 9 | 0 | 28 | 100 | 13 | | 3 | 1 | 3 | 6 | 10 | 1 |
| | 2009 | | 0 | 23 | | | | | | | 1 | 0 | 16 | 76 | 14 | | 0 | 0 | 3 | 4 | 21 | 0 |
| | 2010 | | 9 | 25 | 0 | | | | | | 0 | 0 | 34 | 100 | 8 | 100 | 1 | | 1 | 5 | 3 | 0 |
| Grapes, fresh or dried | 2006 | 89 | 10 | 8 | | | | | | | 7 | 0 | 4 | | 1 | | 0 | 0 | 3 | 0 | | 0 |
| | 2007 | | 13 | 8 | | | 1 | | | | 8 | 1 | 2 | | 1 | | 0 | | 1 | 2 | | 0 |
| | 2008 | | 14 | 9 | 0 | | 1 | | | | 18 | 1 | 2 | | 1 | | 0 | | 1 | 0 | | 0 |
| | 2009 | | 11 | 10 | | | | | | | 0 | 1 | 8 | | 0 | | | | 0 | 1 | | 2 |
| | 2010 | | 13 | 10 | | | 0 | | | | 0 | 0 | 3 | | 0 | | | | 0 | 1 | 1 | 2 |
| Melons, watermelons and papaws (papayas), | 2006 | | 0 | 4 | | | 5 | | | | 16 | | 3 | | 3 | | 0 | 0 | 0 | | | 1 |
| | 2007 | | 0 | 3 | | | 5 | | | | 6 | 1 | 2 | | 1 | | | | 0 | | | 0 |
| | 2008 | | 0 | 5 | | | 3 | | | | 0 | | 2 | | 0 | | 0 | 0 | 0 | 7 | 5 | 0 |
| | 2009 | | 0 | 5 | | | 8 | | | | 0 | 0 | 1 | | 0 | | 0 | | 0 | 6 | 1 | 0 |
| | 2010 | | 1 | 5 | | | 14 | | | | 0 | 0 | 1 | | 0 | | 0 | | 1 | 9 | 0 | 1 |
| Apples, pears and quinces, fresh | 2006 | | 4 | 42 | 66 | | | | | | 14 | 11 | 6 | | 25 | | 31 | 92 | 0 | | 71 | 52 |
| | 2007 | | 6 | 42 | 60 | | | | | | 30 | 6 | 10 | | 28 | | 25 | 97 | 1 | 7 | 56 | 60 |
| | 2008 | | 2 | 39 | 63 | | | | | | 52 | 39 | 7 | | 36 | | 33 | 97 | 0 | 28 | 70 | 71 |
| | 2009 | | 1 | 40 | 72 | | 6 | | | | 86 | 5 | 11 | 24 | 39 | | 17 | 98 | 0 | 17 | 47 | 78 |
| | 2010 | | 5 | 38 | 76 | | | | | | 79 | 0 | 5 | | 49 | | 12 | 100 | 0 | 29 | 93 | 71 |
| Apricots, cherries, peaches (including nectarines), | 2006 | | 4 | 3 | 2 | | | | | | 21 | | 4 | | 2 | | 0 | 8 | 0 | | | 9 |
| | 2007 | | 5 | 3 | 3 | | | | | | 16 | | 3 | | 1 | | | 3 | | | 1 | 12 |
| | 2008 | | 3 | 4 | 2 | | | | | | | | 2 | | 2 | | | 1 | 0 | 7 | 15 | 11 |
| | 2009 | | 3 | 4 | 4 | | | | | | 0 | | 7 | | 1 | | 1 | 1 | | 6 | 24 | 11 |
| | 2010 | | 5 | 4 | 4 | | 13 | | | | 0 | | 5 | | 0 | | | 0 | | 7 | 3 | 11 |
| Other fresh fruit | 2006 | | 1 | 4 | 1 | | | | | | 1 | 33 | 12 | | 9 | | 63 | | 0 | | | 19 |
| | 2007 | | 1 | 3 | 0 | | | | | | 0 | 32 | 10 | | 8 | | 69 | | | 15 | | 15 |
| | 2008 | | 1 | 4 | 0 | | | | | | 0 | 11 | 10 | | 12 | | 58 | 0 | 0 | 3 | | 16 |
| | 2009 | | 1 | 4 | 0 | | | | | | 4 | 14 | 11 | | 15 | | 80 | | 0 | 3 | | 8 |
| | 2010 | | 3 | 4 | 0 | | | | | | 3 | 29 | 11 | | 11 | | 85 | | 0 | 5 | | 13 |
| Fruits, cooked or not, frozen, containing added sugar or sweetener | 2006 | | 0 | 0 | 0 | | | | | | 8 | 18 | 1 | | 25 | | 2 | 0 | | | | 3 |
| | 2007 | | 1 | 0 | 0 | | | | | | 2 | 16 | 1 | | 32 | | 1 | | 2 | | | 0 |
| | 2008 | | 1 | 0 | 0 | 8 | | | | | 2 | 32 | 0 | | 20 | | 2 | | 1 | | | 0 |
| | 2009 | | 0 | 0 | | | | | | | 2 | 4 | 2 | | 11 | | 1 | | 0 | 0 | | |
| | 2010 | | 0 | 0 | | | | | | | | | 2 | | 17 | | | | 2 | 2 | | 0 |
| Fruit provisionally preserved but unsuitable in that state | 2006 | | 0 | | | | | | | | | | 0 | | 0 | | | | | | | |
| | 2007 | | 0 | 0 | | | | | | | | | 0 | | 0 | | 0 | | | | | 0 |
| | 2008 | | 0 | 0 | | | | | | | | | 0 | | | | 0 | 0 | | | | |
| | 2009 | | 0 | 1 | | | | | | | | | 0 | | | | | | | | 0 | |
| | 2010 | | 0 | 0 | | | | | | | | | 0 | | | | | | | | 0 | |
| Dried fruit, mixtures thereof or nuts | 2006 | | 12 | 3 | | | | | | | 0 | 0 | 0 | | 0 | | | 0 | | 1 | | 0 |
| | 2007 | 0 | 13 | 1 | | | 4 | | | | 0 | 0 | 0 | | 0 | | 1 | 0 | | 24 | | 0 |
| | 2008 | | 14 | 1 | | | 1 | | | | 0 | 0 | 0 | | 0 | | 0 | | | 0 | | 0 |
| | 2009 | | 23 | 1 | | | 4 | | | | 0 | 0 | 0 | | 0 | | 0 | | | 0 | | 0 |
| | 2010 | | 15 | 3 | | | 1 | | | | 0 | 0 | 0 | | 0 | | | 0 | | 0 | | 0 |
| Peel of citrus fruits, melons and / or melons, fresh, dried or frozen, etc. | 2006 | | | | | | | | | | | | 0 | | 0 | | | | | | | |
| | 2007 | | | | | | | | | | | | 0 | | | | | | | | | |
| | 2008 | | 0 | | | | | | | | | | 0 | | | | | | | | | |
| | 2009 | | 0 | | | | | | | | | | | | | | 0 | | | | | |
| | 2010 | | | | | | | | | | | | | | | | | | | | | |

From table 3 it is possible to see that Portugal import the majority of the fruits from Spain, some fruits from Germany and France, and some tropical fruits from Brazil and Costa Rica (coconuts, Brazil nuts from Brazil and cashew nuts, fresh or dried, etc, melons, watermelons and papaws (papayas), fresh, bananas, including plantains (platains), fresh or dried and dates, figs, pineapples or pineapple, mangoes, mangosteens, fresh or dried).

**Table 3. Fruits, in different forms, import percentage relatively to the total of each year**

| | Year | South Africa | Angola | Cape Verde | Brazil | Costa Rica | United States of America | China | Turkey | India | Germany | Belgium | Spain | Estonian | France | Greece | Holland | Ireland | Italy | Luxembourg | Poland | United Kingdom |
|---|---|---|---|---|---|---|---|---|---|---|---|---|---|---|---|---|---|---|---|---|---|---|
| Coconuts, Brazil nuts from Brazil and cashew nuts, fresh or dried, | 2006 | | | | 33 | | | | | 8 | 0 | | 6 | | 0 | | 3 | | 0 | | | 0 |
| | 2007 | | | | 30 | 0 | | | | 9 | 1 | | 9 | | 1 | | 4 | | 0 | | | 0 |
| | 2008 | | | | 20 | 0 | 0 | 0 | 17 | 1 | | | 15 | | 2 | | 2 | | 0 | | | 0 |
| | 2009 | | | | 40 | | | 0 | 0 | 3 | 6 | | 8 | | 2 | | 3 | | | | 0 | 3 |
| | 2010 | | | | 42 | 0 | | 0 | | | 3 | | 10 | | 1 | | 4 | | | | | 3 |
| Other nuts, fresh or dried, whether or not shelled or peeled | 2006 | | | | | | 19 | 1 | 1 | 1 | 2 | | 47 | | 10 | 0 | 2 | | 1 | | | 0 |
| | 2007 | | | | 0 | | 16 | 0 | 1 | 1 | 3 | 0 | 48 | | 12 | 0 | 1 | | 1 | | | 0 |
| | 2008 | | | | | | 13 | 0 | 2 | 2 | 4 | 0 | 43 | 0 | 14 | | 1 | | 0 | | | 0 |
| | 2009 | | | | | | 12 | 0 | 1 | 2 | 3 | | 54 | 0 | 17 | | 1 | | 0 | | 0 | 0 |
| | 2010 | | | | | | 12 | 0 | 1 | 2 | 4 | 0 | 48 | 0 | 13 | | 2 | | 0 | | | 0 |
| Bananas, including plantains (platains), fresh or dried | 2006 | | | | 1 | 18 | | | | | 0 | 5 | 23 | | 6 | | 0 | | 7 | | | 0 |
| | 2007 | | | | 1 | 29 | | | | | 0 | 0 | 26 | | 7 | | 0 | | 4 | | | 0 |
| | 2008 | | | | 1 | 23 | 0 | | | | 0 | 0 | 38 | 0 | 5 | | 0 | | 1 | | | 0 |
| | 2009 | | | | 0 | 22 | | | | | 0 | 0 | 27 | | 10 | | 0 | 0 | 0 | | | 1 |
| | 2010 | | | | 0 | 20 | | | | | 0 | 0 | 38 | | 5 | | 0 | | 0 | | | 1 |
| Dates, figs, pineapples or pineapple, mangoes, mangosteens, | 2006 | 0 | | | 23 | 35 | 0 | | 2 | 0 | 0 | 5 | 26 | | 1 | | 3 | | 0 | | | 0 |
| | 2007 | 2 | | | 18 | 31 | 0 | 0 | 2 | 0 | 0 | 4 | 34 | | 1 | | 3 | | 1 | | | 1 |
| | 2008 | 2 | | | 17 | 37 | 0 | | 2 | 0 | 0 | 2 | 33 | 0 | 1 | | 2 | | 0 | | | 0 |
| | 2009 | 1 | | | 14 | 40 | 0 | | 1 | 0 | 0 | 2 | 35 | 0 | 1 | | 1 | 0 | 0 | 1 | | 0 |
| | 2010 | 0 | | | 16 | 45 | 0 | 0 | 1 | 0 | 1 | 1 | 31 | 0 | 1 | | 1 | | 0 | 0 | | 0 |
| Citrus fruit, fresh or dried | 2006 | 3 | | | 9 | | | 0 | | | 10 | 1 | 65 | | 0 | | 1 | | 0 | | | 0 |
| | 2007 | 14 | | | 4 | | | | | | 6 | 0 | 47 | | 1 | | 1 | | 1 | | | 0 |
| | 2008 | 20 | | | 4 | | | | | | 0 | 0 | 45 | 0 | 0 | | 1 | | 4 | | | 1 |
| | 2009 | 12 | | | 6 | | | | | | 1 | 0 | 56 | 0 | 0 | | 2 | | 7 | | | |
| | 2010 | 38 | | | 4 | | | | | | 0 | 0 | 36 | 0 | 0 | | 1 | | 0 | | | |
| Grapes, fresh or dried | 2006 | 11 | | | 0 | | 0 | | 3 | | 4 | 2 | 61 | 0 | 0 | | 1 | | 1 | | | 0 |
| | 2007 | 7 | | | 0 | | | | 3 | | 4 | 0 | 70 | | 0 | 0 | 1 | | 3 | | 0 | 0 |
| | 2008 | 12 | | | | | 0 | | 3 | | 3 | 0 | 63 | 0 | 1 | 0 | 2 | | 5 | | | 1 |
| | 2009 | 7 | | | | | 0 | 0 | 3 | | 1 | 0 | 62 | 0 | 5 | 0 | 1 | | 5 | | | 0 |
| | 2010 | 7 | | | | | 0 | 0 | 2 | | 3 | 0 | 58 | 0 | 3 | | 1 | | 6 | | | 0 |
| Melons, watermelons and papaws (papayas), | 2006 | | | | 23 | 1 | 0 | | | | 2 | | 72 | | 0 | | 0 | | 0 | | | |
| | 2007 | | | | 21 | 1 | 0 | | | | 3 | | 73 | 0 | 0 | 0 | | | | | | 0 |
| | 2008 | | | | 23 | 0 | 0 | | | | 2 | | 73 | 0 | 0 | 0 | 1 | | | | | 0 |
| | 2009 | 0 | | | 24 | 1 | | | | | 2 | | 71 | 0 | 0 | 0 | 0 | | | 1 | | |
| | 2010 | | | | 25 | 1 | | | | | 3 | | 70 | 0 | 0 | 0 | 0 | 0 | 0 | | | 0 |
| Apples, pears and quinces, fresh | 2006 | 2 | | | 3 | | | | | | 13 | 0 | 31 | | 20 | | 2 | | 1 | | | |
| | 2007 | 4 | | | 7 | | | 0 | | | 9 | 0 | 29 | | 28 | | 2 | | 3 | | | 0 |
| | 2008 | 6 | | | 6 | | | 1 | | | 2 | 0 | 33 | 0 | 26 | | 1 | | 5 | | 0 | 0 |
| | 2009 | 8 | | | 8 | | | 0 | | | 1 | 0 | 31 | 0 | 18 | | 2 | | 10 | | 0 | 0 |
| | 2010 | 5 | | | 10 | | | 0 | | | 0 | 0 | 29 | 0 | 27 | | 1 | | 7 | | | 1 |
| Apricots, cherries, peaches (including nectarines), | 2006 | 2 | | | | | | | | | 7 | 0 | 88 | | 1 | | 0 | | 0 | | | |
| | 2007 | 2 | | | | | | | | | 8 | 0 | 87 | | 1 | | 0 | | | | | |
| | 2008 | 7 | | | | | | | | | 2 | 0 | 89 | 0 | 0 | | 0 | | 0 | | | 0 |
| | 2009 | 4 | | | | | | | 0 | | 2 | 1 | 92 | 0 | 0 | | 0 | | 0 | | | |
| | 2010 | 1 | | | | | | | | | 1 | 1 | 96 | 0 | 0 | 0 | 0 | | 0 | | | 0 |
| Other fresh fruit | 2006 | | | | 0 | | | | | 0 | 5 | 2 | 72 | | 2 | | 8 | | 5 | | | 0 |
| | 2007 | | | | 1 | | 0 | | | 0 | 3 | 3 | 73 | 0 | 2 | | 6 | | 5 | | | 0 |
| | 2008 | 0 | | | 0 | | | 0 | | 0 | 1 | 2 | 75 | 0 | 3 | 0 | 6 | | 7 | | | 0 |
| | 2009 | 0 | | | 1 | 0 | | | | | 2 | 2 | 65 | 0 | 11 | 0 | 6 | | 7 | | | 0 |
| | 2010 | 0 | | | 1 | | | | | 0 | 3 | 7 | 72 | 0 | 4 | | 5 | | 4 | | | 0 |
| Fruits, cooked or not, frozen, containing added sugar or sweetener | 2006 | | | | | | | 1 | 1 | | 2 | 14 | 22 | | 16 | 6 | 11 | | | | 1 | 0 |
| | 2007 | | | | | | | | 3 | | 1 | 26 | 29 | | 12 | 9 | 12 | | 0 | | 1 | 0 |
| | 2008 | | | | | | | 0 | 1 | | 1 | 6 | 17 | | 5 | 9 | 30 | | 4 | | | 1 |
| | 2009 | | | | | | | 2 | 2 | | 1 | 15 | 25 | | 13 | 15 | 17 | | 4 | | | 0 |
| | 2010 | | | | 8 | | 0 | | 4 | | 1 | 13 | 18 | | 18 | 14 | 11 | | 5 | 0 | | 0 |
| Fruit provisionally preserved but unsuitable in that state | 2006 | | | | | | | | | | 22 | 0 | 73 | | 4 | | | | 1 | | | 0 |
| | 2007 | | | | | | | | | | 20 | | 72 | | 7 | | | | 1 | | | 0 |
| | 2008 | | | | | | | | | | 4 | 0 | 90 | | 6 | | | | 0 | | | 0 |
| | 2009 | | | | | | | 0 | | | 12 | | 80 | | 6 | | | | 1 | | | |
| | 2010 | | | | | | | | | | 13 | | 78 | | 8 | | | | 1 | | | |
| Dried fruit, mixtures thereof or nuts | 2006 | 2 | | | 0 | | 0 | 1 | 12 | 0 | 12 | | 42 | | 14 | | 1 | | 0 | | | 0 |
| | 2007 | 0 | | | 1 | | 1 | 1 | 14 | 0 | 14 | 0 | 38 | | 13 | | 1 | | 1 | | | 0 |
| | 2008 | 0 | | | 3 | | 0 | 1 | 16 | | 12 | 1 | 42 | | 10 | | 1 | | 1 | | | 1 |
| | 2009 | 1 | | | 1 | | 0 | 5 | 11 | 0 | 15 | | 36 | | 3 | | 1 | | 0 | | 0 | 1 |
| | 2010 | | | | 0 | | 0 | 3 | 14 | | 24 | | 33 | | 3 | | 1 | | 0 | | | 1 |
| Peel of citrus fruits, melons and / or melons, fresh, dried or frozen, etc. | 2006 | | | | | | | | | | 0 | | 100 | | | | | | | | | |
| | 2007 | | | | | | | | | | 88 | 12 | | | | | | | | | | |
| | 2008 | | | | | 72 | 0 | | | | 2 | 1 | 2 | | 11 | | | | | | | 3 |
| | 2009 | | | | | | | | | | 0 | | 60 | | 0 | | | | 40 | | | |
| | 2010 | | | | | | | | | | | 38 | 62 | | | | | | | | | |

Portugal export the majority of the fruits to Spain (table 4) and some fruits to France (namely, fruits, cooked or not, frozen, containing added sugar or sweetener), to Italy (bananas, including plantains (platains), fresh or dried and dates, figs, pineapples or pineapple, mangoes, mangosteens, fresh or dried), to the United Kingdom, to Angola (dried fruit, mixtures thereof or nuts) and to Cape Verde.

**Table 4. Fruits, in different forms, export percentage relatively to the total of each year**

| Product | Year | South Africa | Angola | Cape Verde | Brazil | Costa Rica | United States of America | China | Turkey | India | Germany | Belgium | Spain | Estonian | France | Greece | Holland | Ireland | Italy | Luxembourg | Poland | United Kingdom |
|---|---|---|---|---|---|---|---|---|---|---|---|---|---|---|---|---|---|---|---|---|---|---|
| Coconuts, Brazil nuts from Brazil and cashew nuts, fresh or dried, | 2006 | | 62 | 26 | | | 2 | | | | 0 | 0 | 0 | | 4 | | 0 | 0 | 0 | 0 | | 0 |
| | 2007 | | 32 | 57 | | | 4 | | | | 0 | | 0 | | 0 | | 0 | 0 | 0 | 0 | | 0 |
| | 2008 | 0 | 41 | 41 | | | 3 | | 0 | | 0 | 1 | | 0 | | 0 | 0 | 0 | 0 | | 1 |
| | 2009 | 0 | 31 | 19 | | | 3 | | | | 0 | 0 | 43 | | 1 | | 0 | | 0 | 0 | | 0 |
| | 2010 | | 55 | 17 | | | 4 | 0 | | | 0 | 0 | 1 | | 1 | | 0 | 0 | 0 | 0 | | 14 |
| Other nuts, fresh or dried, whether or not shelled or peeled | 2006 | 0 | 5 | 0 | 13 | | 1 | | 0 | 2 | 0 | | 47 | | 11 | | 0 | 0 | 14 | 1 | | 3 |
| | 2007 | 0 | 5 | 0 | 17 | | 1 | | 1 | 2 | 1 | 2 | 43 | | 11 | 1 | 0 | 0 | 14 | 0 | 0 | 3 |
| | 2008 | 0 | 7 | 0 | 13 | | 1 | | | 0 | 1 | 1 | 46 | | 21 | | 0 | 0 | 6 | 1 | | 1 |
| | 2009 | 0 | 5 | 0 | 13 | | 1 | | | 1 | 1 | 4 | 32 | | 20 | | 0 | | 17 | 3 | | 1 |
| | 2010 | 0 | 4 | 0 | 14 | | 1 | | | | 1 | 2 | 40 | | 12 | | 1 | 0 | 18 | 1 | | 2 |
| Bananas, including plantains (platains), fresh or dried | 2006 | | 0 | 0 | | | | | | | 0 | 0 | 26 | | 21 | | | 0 | 42 | | | 10 |
| | 2007 | | 0 | 0 | | | | | | | 0 | | 41 | | 15 | | | | 32 | | 0 | 11 |
| | 2008 | | 0 | 0 | | | | | | | 0 | | 50 | | | | | | 50 | 0 | | 0 |
| | 2009 | | | 0 | | | | | | | | | 35 | | 0 | | | | 65 | 0 | | |
| | 2010 | | 0 | 0 | | | | | | | | | 48 | | 0 | | 0 | | 52 | 1 | | |
| Dates, figs, pineapples or pineapple, mangoes, mangosteens, | 2006 | | 1 | 1 | | | 0 | | | | 1 | 0 | 76 | | 1 | | 1 | 0 | 18 | 0 | | 0 |
| | 2007 | 2 | 2 | 2 | | | 0 | | | | 0 | 0 | 64 | | 2 | | 3 | | 25 | 0 | | 0 |
| | 2008 | 1 | 2 | 1 | | | 0 | | 0 | | 1 | 0 | 57 | 0 | 3 | 0 | 2 | 0 | 30 | 1 | | 0 |
| | 2009 | 0 | 2 | 1 | | | 0 | | | | 0 | 0 | 57 | | 1 | 0 | 1 | 0 | 37 | 0 | 0 | 0 |
| | 2010 | 0 | 1 | 1 | | | 0 | | | | 0 | 0 | 57 | | 0 | | 1 | | 39 | 0 | | 0 |
| Citrus fruit, fresh or dried | 2006 | | 0 | 5 | | | | | | | 0 | | 71 | | 14 | | | | 3 | 0 | 3 | 3 |
| | 2007 | | 0 | 4 | | | | | | | 0 | 1 | 72 | | 13 | | 0 | | 2 | | 4 | 1 |
| | 2008 | | 0 | 3 | | | | | | | 1 | 0 | 71 | 0 | 16 | | 1 | 0 | 2 | 0 | 1 | 1 |
| | 2009 | | 0 | 5 | | | | | | | 0 | 0 | 59 | 0 | 24 | | 0 | 0 | 5 | 0 | 1 | 0 |
| | 2010 | | 1 | 3 | | 0 | | | | | 0 | 0 | 88 | 0 | 6 | 0 | 0 | | 0 | 0 | 0 | 0 |
| Grapes, fresh or dried | 2006 | 1 | 8 | 6 | | | | | | | 2 | | 57 | | 5 | | 1 | 0 | 17 | 0 | | 1 |
| | 2007 | | 15 | 11 | | | 0 | | | | 2 | 1 | 42 | | 19 | | 0 | | 5 | 0 | | 4 |
| | 2008 | | 14 | 9 | 0 | | 0 | | | | 11 | 1 | 45 | | 6 | | 1 | | 6 | 0 | | 1 |
| | 2009 | | 6 | 6 | | | | | | | 0 | 0 | 80 | | 1 | | | | 1 | 0 | | 6 |
| | 2010 | | 10 | 9 | | | 0 | | | | 0 | 0 | 64 | | 1 | | | | 3 | 0 | 0 | 9 |
| Melons, watermelons and papaws (papayas), | 2006 | | 0 | 3 | | | 0 | | | | 4 | | 45 | | 27 | | 0 | 0 | | 0 | | 10 |
| | 2007 | | 1 | 5 | | | 1 | | | | 2 | 1 | 60 | | 21 | | | | 1 | | | 3 |
| | 2008 | | 0 | 9 | | | 0 | | | | 0 | | 72 | | 3 | | 2 | 0 | 1 | 1 | 4 | 2 |
| | 2009 | | 1 | 12 | | | 2 | | | | 1 | 0 | 58 | | 7 | | 4 | | 3 | 5 | 1 | 2 |
| | 2010 | | 1 | 9 | | | 3 | | | | 0 | 0 | 62 | | 1 | | 0 | | 7 | 4 | 1 | 5 |
| Apples, pears and quinces, fresh | 2006 | | 0 | 2 | 17 | | | | | | 0 | 0 | 8 | | 16 | | 5 | 8 | 0 | | 2 | 29 |
| | 2007 | | 0 | 3 | 15 | | | | | | 0 | 0 | 11 | | 17 | | 5 | 8 | 0 | 0 | 2 | 32 |
| | 2008 | | 0 | 2 | 10 | | | | | | 2 | 1 | 8 | | 20 | | 7 | 9 | 0 | 1 | 3 | 32 |
| | 2009 | | 0 | 2 | 18 | | 0 | | | | 4 | 0 | 11 | 0 | 19 | | 4 | 7 | 0 | 0 | 1 | 28 |
| | 2010 | | 0 | 2 | 27 | | | | | | 3 | 0 | 7 | | 20 | | 3 | 6 | 0 | 0 | 4 | 22 |
| Apricots, cherries, peaches (including nectarines), | 2006 | | 2 | 2 | 4 | | | | | | 3 | | 35 | | 9 | | 0 | 6 | 0 | | | 38 |
| | 2007 | | 2 | 2 | 5 | | | | | | 2 | | 32 | | 6 | | | 2 | | | 0 | 50 |
| | 2008 | | 1 | 2 | 3 | | | | | | | | 25 | | 8 | | | 1 | 0 | 1 | 5 | 47 |
| | 2009 | | 1 | 2 | 7 | | | | | | 0 | | 51 | | 3 | | 1 | 1 | | 1 | 4 | 29 |
| | 2010 | | 2 | 2 | 10 | | 1 | | | | 0 | | 54 | | 1 | | | 0 | | 1 | 1 | 25 |
| Other fresh fruit | 2006 | | 0 | 1 | 0 | | | | | | 0 | 3 | 34 | | 14 | | 25 | | 0 | | | 23 |
| | 2007 | | 0 | 0 | 0 | | | | | | 0 | 3 | 28 | | 13 | | 34 | | | 0 | | 20 |
| | 2008 | | 0 | 1 | 0 | | | | | | 0 | | 29 | | 17 | | 29 | 0 | 0 | 0 | | 18 |
| | 2009 | | 0 | 1 | 0 | | | | | | 0 | 1 | 28 | | 18 | | 46 | | 0 | 0 | | 7 |
| | 2010 | | 0 | 0 | 0 | | | | | | 0 | 1 | 32 | | 9 | | 48 | | 0 | 0 | | 8 |
| Fruits, cooked or not, frozen, containing added sugar or sweetener | 2006 | | 0 | 0 | 0 | | | | | | 1 | 3 | 4 | | 75 | | 1 | 0 | | | | 8 |
| | 2007 | | 0 | 0 | 0 | | | | | | 0 | 3 | 3 | | 86 | | 1 | | 3 | | | 0 |
| | 2008 | | 0 | 0 | 0 | | 0 | | | | 1 | 7 | 2 | | 83 | | 2 | | 2 | | | 0 |
| | 2009 | | 0 | 0 | | | | | | | 1 | 1 | 24 | | 63 | | 2 | | 2 | 0 | | |
| | 2010 | | 0 | 0 | | | | | | | | 0 | 21 | | 65 | | | | 6 | 0 | | 0 |
| Fruit provisionally preserved but unsuitable in that state | 2006 | | 8 | | | | | | | | | | | | 88 | | 0 | | | | | |
| | 2007 | | 0 | 1 | | | | | | | | | 34 | | 65 | | 0 | | | | | 0 |
| | 2008 | | 34 | 42 | | | | | | | | | 3 | | | | 17 | 5 | | | | |
| | 2009 | | 6 | 36 | | | | | | | | | 55 | | | | | | | | 3 | |
| | 2010 | | 4 | 28 | | | | | | | | | 60 | | | | | | | | 4 | |
| Dried fruit, mixtures thereof or nuts | 2006 | | 73 | 16 | | | 0 | | 0 | 0 | 0 | | 3 | | | | 0 | | 1 | | | 0 |
| | 2007 | 0 | 52 | 6 | | | 1 | | 0 | 0 | 0 | | 9 | | 2 | | 17 | 0 | 0 | | | 0 |
| | 2008 | | 62 | 5 | | | 1 | | 0 | 0 | 0 | | 11 | | 4 | | 0 | | 1 | | | 0 |
| | 2009 | | 81 | 5 | | | 2 | | 0 | 0 | 0 | | 7 | | 1 | | 0 | | 0 | | | 0 |
| | 2010 | | 69 | 14 | | | 1 | | 0 | 0 | 0 | | 10 | | 1 | | | 0 | 0 | | | 0 |
| Peel of citrus fruits, melons and / or melons, fresh, dried or frozen, etc. | 2006 | | | | | | | | | | | | 100 | | | | | | | | | |
| | 2007 | | | | | | | | | | | | 100 | | | | | | | | | |
| | 2008 | | 3 | | | | | | | | | | 97 | | | | | | | | | |
| | 2009 | | 2 | | | | | | | | | | | | | | 98 | | | | | |
| | 2010 | | | | | | | | | | | | | | | | | | | | | |

## 3. Estimations results for the neoclassical model with panel data and volatility analysis

They were made several estimations, based in the absolute convergence model, of Solow (1956), with panel data, following procedures of Islam (1995), using econometric methods, in the informatics program Stata, like fixed effects, random effects and dynamic panel data. The estimations were made with data from 2006 to 2010, for the different countries with international trade of fruits with Portugal and for different forms of fruits. There were made another estimations with the data in percentage, like are offered in the table 1, 2, 3 and 4 presented in the previous section of this work.

All the results show that there is not statistically significance for the Portuguese international trade of fruits.

These results with the lack of stationary of the data verified in the volatility analysis, show that there is not an objective policy for the international trade of fruits in Portugal and consequently there is not a policy for the Portuguese fruit production. Like the Keynesian theory say, the export is the engine of the output of each sector.

So, in light of is the common agricultural policy, Portugal must do an adjusted national agricultural policy for the fruit sector.

**Table 5. Results from the absolute convergence model for all fruits import (absolute values)**

|  | Const.[1] | Coef.[2] | F/Wald(mod.)[3] | F(Fe_OLS)[4] | Corr(u_i)[5] | F(Re_OLS)[6] | Hausman[7] | $R^2$[8] | N.O.[9] | N.I.[10] |
|---|---|---|---|---|---|---|---|---|---|---|
| FE[11] | 13.102* (22.170) | -1.047* (-22.200) | 492.870* | 4.140* | -0.914 | ------- | ------- | 0.585 | 497 | ------- |
| RE[12] | 2.816* (7.810) | -0.227* (-8.000) | 63.950* | ------- | ------- | 11.250* | 473.900* | 0.585 | 497 | ------- |
| OLS | ------- | ------- | ------- | ------- | ------- | ------- | ------- | ------- | ------- | ------- |
| DPD[13] | 19.476* (18.630) | -1.533* (-18.680) | 417.730* | ------- | ------- | ------- | ------- | ------- | 220 | 5 |

Note: 1, Constant; 2, Coefficient; 3, Test F for fixed effects model and test Wald for random effects and dynamic panel data models; 4, Test F for fixed effects or OLS (Ho is OLS); 5, Correlation between errors and regressors in fixed effects; 6, Test F for random effects or OLS (Ho is OLS); 7, Hausman test (Ho is GLS); 8, R square; 9, Number of observations; 10, Number of instruments;, 11, Fixed effects model; 12, Random effects model; 13, Dynamic panel data model; *, Statically significant at 5%.

**Table 6. Results from the absolute convergence model for all fruits export (absolute values)**

|  | Const.[1] | Coef.[2] | F/Wald(mod.)[3] | F(Fe_OLS)[4] | Corr(u_i)[5] | F(Re_OLS)[6] | Hausman[7] | $R^2$[8] | N.O.[9] | N.I.[10] |
|---|---|---|---|---|---|---|---|---|---|---|
| FE[11] | 10.459* (17.390) | -0.937* (-17.450) | 304.350* | 3.250* | -0.895 | ------- | ------- | 0.462 | 505 | ------- |
| RE[12] | 2.619* (7.430) | -0.238* (-7.800) | 60.800* | ------- | ------- | 2.810* | 250.250* | 0.462 | 505 | ------- |
| OLS | ------- | ------- | ------- | ------- | ------- | ------- | ------- | ------- | ------- | ------- |
| DPD[13] | 16.420* (12.730) | -1.439* (-12.690) | 263.040* | ------- | ------- | ------- | ------- | ------- | 217 | 5 |

**Table 7. Results from the absolute convergence model for all fruits import (percentage values relatively to the total of each country)**

|  | Const.[1] | Coef.[2] | F/Wald(mod.)[3] | F(Fe_OLS)[4] | Corr(u_i)[5] | F(Re_OLS)[6] | Hausman[7] | $R^2$[8] | N.O.[9] | N.I.[10] |
|---|---|---|---|---|---|---|---|---|---|---|
| FE[11] | 1.226* (16.780) | -1.079* (-22.060) | 486.450* | 3.800* | -0.883 | ------- | ------- | 0.582 | 497 | ------- |
| RE[12] | 0.282* (3.290) | -0.273* (-8.750) | 76.580* | ------- | ------- | 9.520* | 457.140* | 0.582 | 497 | ------- |
| OLS | ------- | ------- | ------- | ------- | ------- | ------- | ------- | ------- | ------- | ------- |
| DPD[13] | 1.867* (15.700) | -1.549* (-17.380) | 377.090* | ------- | ------- | ------- | ------- | ------- | 220 | 5 |

**Table 8. Results from the absolute convergence model for all fruits export (percentage values relatively to the total of each country)**

|  | Const.[1] | Coef.[2] | F/Wald(mod.)[3] | F(Fe_OLS)[4] | Corr(u_i)[5] | F(Re_OLS)[6] | Hausman[7] | $R^2$[8] | N.O.[9] | N.I.[10] |
|---|---|---|---|---|---|---|---|---|---|---|
| FE[11] | 0.154* (2.460) | -0.910* (-16.970) | 287.950* | 3.040* | -0.878 | ------- | ------- | 0.449 | 505 | ------- |
| RE[12] | -0.108 (-1.020) | -0.251* (-7.980) | 63.680* | ------- | ------- | 3.680* | 230.210* | 0.449 | 505 | ------- |
| OLS | ------- | ------- | ------- | ------- | ------- | ------- | ------- | ------- | ------- | ------- |
| DPD[13] | 0.498* (6.020) | -1.440* (-12.140) | 239.650* | ------- | ------- | ------- | ------- | ------- | 217 | 5 |

**Table 9. Results from the absolute convergence model for all fruits import (percentage values relatively to the total of each year)**

|  | Const.[1] | Coef.[2] | F/Wald(mod.)[3] | F(Fe_OLS)[4] | Corr(u_i)[5] | F(Re_OLS)[6] | Hausman[7] | $R^2$ [8] | N.O.[9] | N.I.[10] |
|---|---|---|---|---|---|---|---|---|---|---|
| FE[11] | 0.161* (3.330) | -1.059* (22.210) | 493.450* | 4.220* | -0.897 | ------- | ------- | 0.585 | 497 | ------- |
| RE[12] | -0.046 (-0.490) | -0.293* (-9.150) | 83.670* | ------- | ------- | 7.980* | 470.680* | 0.585 | 497 | ------- |
| OLS | ------- | ------- | ------- | ------- | ------- | ------- | ------- | ------- | ------- | ------- |
| DPD[13] | 0.467* (8.740) | -1.568* (-18.700) | 425.600* | ------- | ------- | ------- | ------- | ------- | 220 | 5 |

**Table 10. Results from the absolute convergence model for all fruits export (percentage values relatively to the total of each year)**

|  | Const.[1] | Coef.[2] | F/Wald(mod.)[3] | F(Fe_OLS)[4] | Corr(u_i)[5] | F(Re_OLS)[6] | Hausman[7] | $R^2$ [8] | N.O.[9] | N.I.[10] |
|---|---|---|---|---|---|---|---|---|---|---|
| FE[11] | -0.025 (-0.400) | -0.923* (-16.330) | 266.670* | 2.900* | -0.863 | ------- | ------- | 0.430 | 505 | ------- |
| RE[12] | -0.131 (-1.260) | -0.257* (-7.660) | 58.670* | ------- | ------- | 2.280 | 214.390* | 0.430 | 505 | ------- |
| OLS | ------- | ------- | ------- | ------- | ------- | ------- | ------- | ------- | ------- | ------- |
| DPD[13] | 0.218* (3.230) | -1.397* (-11.660) | 239.530* | ------- | ------- | ------- | ------- | ------- | 217 | 5 |

## 4. Conclusions

The Europe, namely Spain, is the principal partner of Portugal to the international trade of fruits. What is expected, because the cost of transport. The transport of fruits is not cheap and transport these products in long distance worse.

This is in line with of the new economic geography what says that the transport costs are important and the economic sectors have a tendency to be close to minimizing the cost of transportation.

Portugal needs a new national policy to fruit sector, not only to the international trade, but also to the production. Is not easy to formulate a new national policy, because the limitations of the common agricultural policy from the European Union, but the Portuguese authorities must be able to find new ways for the sector in line with the European policies.

## 5. References

Bunte, F. (2005). *Liberalising EU Imports for Fruits and Vegetables*. Paper prepared for presentation at the XIth Congress of the EAAE (European Association of Agricultural Economists), 'The Future of Rural Europe in the Global Agri-Food System', Copenhagen, Denmark, August 24-27, 2005.

Cioffi, A. and dell´Aquila, C. (2004). *The effects of trade policies for fresh fruit and vegetables of the European Union*. Food Policy, Volume 29, Issue 2, 169–185.

Cioffi A.; Santeramo F.G.; and Vitale C.D. (2010). *The Price Stabilisation Effects of the EU entry price scheme for fruits and vegetables.* MPRA Paper No. 24828.

Coque, J.M.G.A. and Selva, M.L.M. (2007). *A Gravity Approach to Assess the Effects of Association Agreements on Euromediterranean Trade of Fruits and Vegetables*. MPRA Paper No. 4124.

Emmy, F.A. and Ismail, M.M. (2009). *Trade Performance of Fruit and Vegetable Industry in Selected ASEAN Countries*. MPRA Paper No. 16928.

Fragoso, R.M.S.; Marques, C.A.F.; Lucas, M.R.V.; Martins, M.B.; and Jorge, R.F. (2009). *THE ECONOMIC EFFECTS OF COMMON AGRICULTURAL POLICY TRENDS ON MONTADO ECOSYSTEM IN SOUTHERN PORTUGAL*. CEFAGE-UE Working Paper 2009/12.

Islam, N. (1995). *Growth Empirics : A Panel Data Approach.* Quarterly Journal of Economics, 110, 1127-1170.

Karemera, D.; Sykes, V.D.; and Reuben, L.J. (2007). *Trade Creation, Trade Diversion effects of NAFTA on vegetable and fruit trade flows*. World Review of Entrepreneurship, Management and Sustainable Development, Volume 3, Nº 2, 142-157.

Seck, A.; Cissokho, L.; Makpayo, K.; and Haughton, J. (2010). *How important are non-tariff barriers to agricultural trade within ECOWAS?*. Department of Economics, Suffolk University, Research Working Papers Nº 2010-3.

Solow, R. (1956). *A Contribution to the Theory of Economic Growth.* Quarterly Journal of Economics.